# Alignment-Tolerant Optical Fi-Wi-Fi Bridge Assisted by a Focal Plane Array Beamformer as Air Interface

Florian Honz, and Bernhard Schrenk
*AIT Austrian Institute of Technology, Center for Digital Safety & Security / Security & Communication Technologies, 1210 Vienna, Austria.*
*Author e-mail address: florian.honz@ait.ac.at*

## ABSTRACT

**Terrestrial free-space optical (FSO) links are an ideal candidate to extend the bandwidth continuum offered by fiber networks, yet at the expense of unfavorable cost credentials due to highly complex opto-mechanical setups. As a response to this challenge, we present a simple fiber-based focal plane array (FPA) architecture which contributes beamforming functionality to an FSO link that bridges the gap between two single-mode fiber ports. Through use of space-switched and wavelength-routed beamforming networks, which together with a photonic lantern provide a compact arrangement of up to 61 fine-pitched optical antenna elements on a fiber tip, we experimentally demonstrate that our FPA can assist the channel optimization of an FSO link after rough initial beam pointing. We show that favorable coupling conditions can be accomplished between two standard single-mode fibers, which enables us to successfully retain the fiber continuum by achieving error-free 10 Gb/s/λ data transmission in a space-switched out-door fiber-wireless-fiber scenario.**

## INTRODUCTION

The push of network bandwidth and data processing closer to the edge of the communication infrastructure makes the delivery of fiber-grade capacity a prime objective. However, the deployment of fiber close to the network users often comes with high expenditures, whereas the provisioning of communication technology in access and last-mile segments calls for cost-efficient solutions. Optical wireless communication and, in particular, free-space optical (FSO) links are considered as a practical tool to bridge bandwidth gaps in the overall network infrastructure [1, 2]. Towards this, first practical deployments of FSO links have demonstrated sustainable economic models [3]. However, many challenges exist when extending the bandwidth continuum of wired fiber networks over the air, such as a highly efficient optical coupling between FSO transceivers [4], the mitigation of atmospheric propagation effects [5], robust transmission under poor weather conditions [6], or the FSO link termination through single-mode fiber at the remote link end, to eventually enable coherent transmission over FSO communications [2, 7]. Although impressive accomplishments have been made by showing capacities of more than 14 Tb/s per link [8] or 1.1 Tb/s/λ [9], these implementations come at the expense of complex system setups. In contrast to that, short-reach terrestrial FSO links are required to strike a compromise in terms of system complexity while still aiming at fiber-grade transmission capacity. Optical beam steering and -forming are considered as enablers for simplified FSO links. While beam steering can be supported by adopting fast-tunable optics such as micro-mirrors [7, 10], multi-element beamformers allow for a spatial processing of emitted optical communication beams for the purpose of beam steering or turbulence mitigation through coherent beam combining [11]. Towards this direction, integrated optical multi-element antenna configurations that can support a migration towards cost-effective FSO communication systems have been demonstrated, including optical phased arrays (OPA) [12-15] and focal plane arrays (FPA) [16, 17].

This work extends our initial investigation [18] on a fiber-wireless-fiber (Fi-Wi-Fi) bridge that aims at fiber continuity within the FSO system till the final stage of beam collimation: By involving established fiber-optic components that have proven their reliability in the field of fixed fiber-optic communication networks, we are able to implement a 61-element FPA beamformer for alignment-tolerant FSO links. We will evaluate space-switched as well as wavelength-routed beamformer networks and benchmark the performance of selected FPA designs for an out-door Fi-Wi-Fi bridge installed as a 63-m roof-top FSO link, demonstrating that the FPAs involved at the transmitting and receiving FSO terminals are able to establish a good coupling efficiency after no more than coarse initial beam pointing, to eventually enable error-free data transmission at 10 Gb/s/λ.

This manuscript is organized as follows. Section II introduces the space-switched and wavelength-routed FPA architectures and characterizes their constituent elements. Section III introduces the experimental framework that has been used to evaluate both beamformers in FSO links. Section IV then discusses the performance of FSO links that build on the space-switched FPA, while Section V elaborates on the wavelength-routed FPA and its contribution to simultaneous channel sounding. Finally, Section VI concludes the work.

## AIR INTERFACE BASED ON FOCAL PLANE ARRAY ON A FIBER TIP

A FPA-based beamformer features simple control and does not necessitate exhaustive calibration, as it is for example

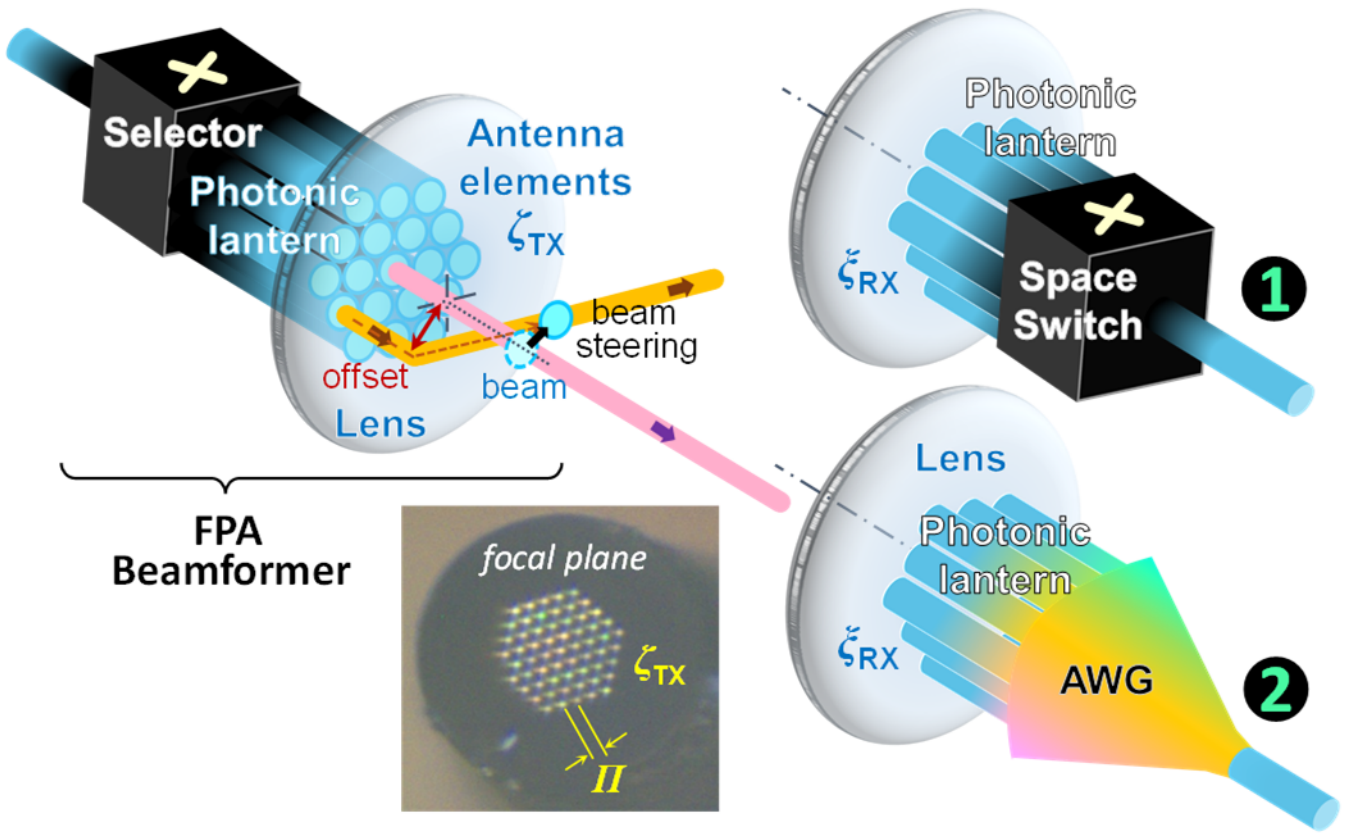

**Fig. 1.** FPA beamformer implementations based on a space-switched (**1**) and wavelength-routed (**2**) beamforming network. The inset presents a photograph of the end-facet of the photonic lantern at the focal plane of the beamformer.

TABLE I

AIR INTERFACE CONFIGURATIONS AT THE RECEIVING SITE.

| Characteristic \ Beamforming | | ❶ Space-switched | ❷ λ-routed |
|---|---|---|---|
| FPA Architecture | number of antenna elements | $\xi_{RX}$ = 61 | $\xi_{RX}$ = 61 |
| | antenna element pitch | 37 µm | 37 µm |
| | fill factor | 5.4% | 5.4% |
| Beamforming network | Architecture | 1×32 or 1×64 MEMS switch | 96-channel 50-GHz AWG |
| | Spectral transparency | wide: O- to L-band | narrow: 50-GHz drop, FSR of 5.96 THz |
| **Link quality monitoring** | | directly on ROP or time-interleaved wideband sounding | simultaneous channel sounding on adjacent FSR |

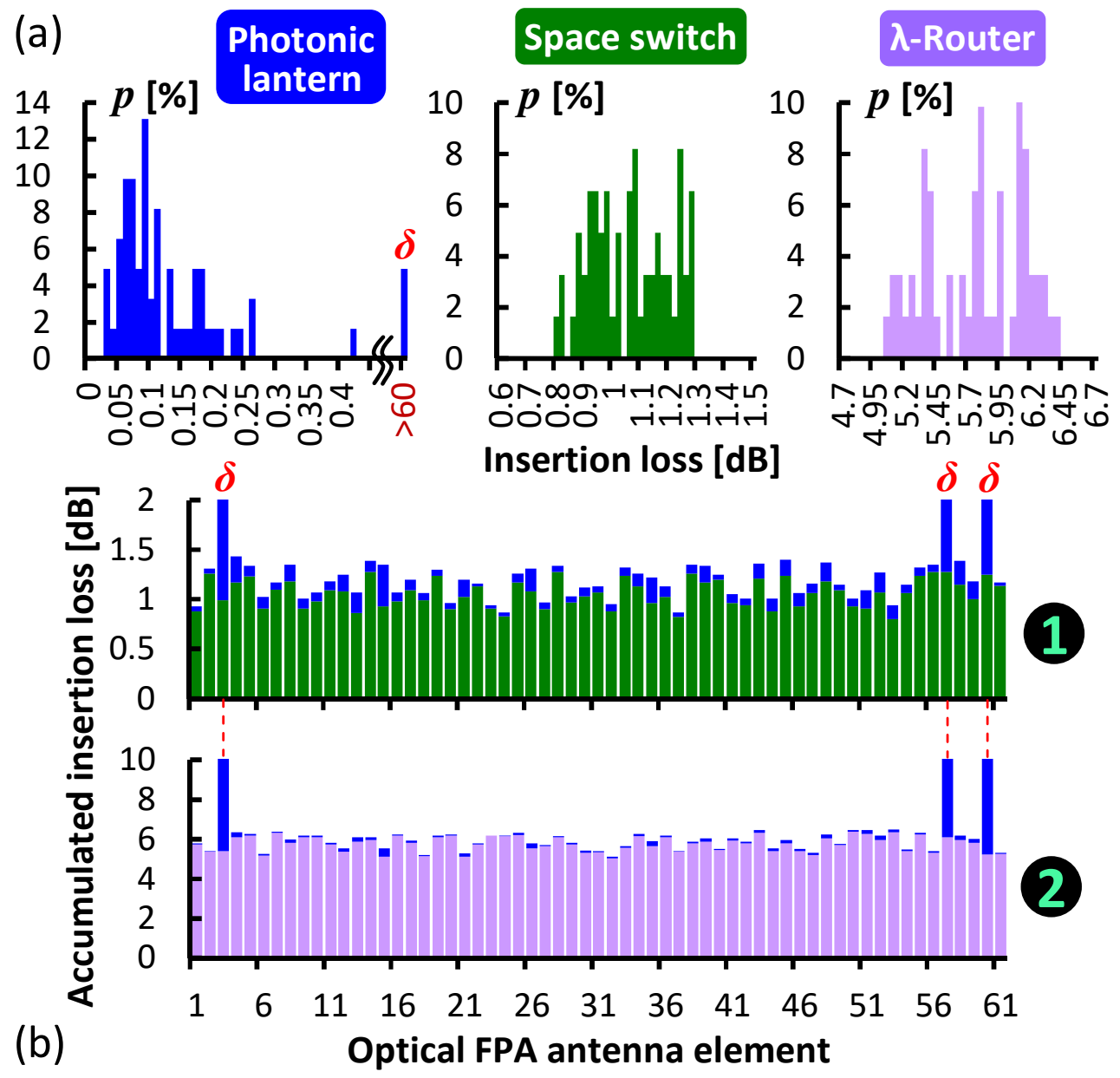

**Fig. 2.** (a) Insertion loss for all constituent elements of a 61-channel FPA beamformer and (b) accumulated loss for all antenna elements of a space-switched (**1**) and wavelength-routed (**2**) FPA configuration.

required to retrieve the intrinsic phase map of OPA-based beamformers [12, 13]. Moreover, FPAs can be realized with a larger pitch between its antenna elements, which can ease the practical implementation and fabrication in case that short-wavelength operation is considered. While FPA beamformers have been demonstrated as chip-scale solutions [16, 17], the present investigation employs a FPA at the tip of a fiber. Figure 1 presents this fiber-based FPA concept. The FPA provides the air interface of a point-to-point FSO link that serves as an optical bridge between two ITU-T G.652B-compatible single-mode fibers (SMF). A colorless space switch (**1**) and a wavelength-router (**2**) are considered as two options for the beamforming network that completes the FPA-enabled beamformer. Table 1 summarizes the characteristics for both implementations.

The focal plane of the beamformer features 61 individual single-mode cores of a photonic lantern with an identical numerical aperture (NA) of 0.12 for each core. These cores at the end-facet of the lantern are arranged in a hexagonal pattern and used as antenna elements, as show in the inset of Fig. 1. The pitch $\Pi$ between the elements was 37 µm and the fill factor for all cores over the focal plane is 5.4% or -12.7 dB. The overall ensemble of FPA antenna elements is centered to the optical axis of a lens with focal length $f$. According to the notion of a FPA beamformer, the offset between the focal center of the lens and the actually illuminated core determines the steering angle of the light beam that is emitted by the FPA. Due to the rotation symmetry of the FPA, the maximum number of $N$ = 9 antenna elements in a row of the hexagonal arrangement then yields the field-of-view of the beamformer according to

$$FoV = \frac{N \cdot \Pi}{f} \quad (1)$$

The discretization $\upsilon$ when steering the projected beam at a link distance $L$ is given by

$$\upsilon = L \tan\left(\frac{\Pi}{f}\right) \quad (2)$$

For a fixed design of the photonic lantern, the optical lens can then be chosen according to the operational parameters required for a specific FSO link setting. In general, the field-of-view is enhanced compared to a single fiber core, which enables us to establish optical coupling between two SMF even for unfavourable initial alignment conditions. To do so, a selection of the antenna element $\zeta_{TX}$ and $\xi_{RX}$ at the transmitting and receiving FPA beamformers is facilitated through the respective beamforming networks. While we will exclusively use a space-switched architecture for the purpose of antenna element selection at the transmitter, we will evaluate space-switched (**1**) and wavelength-routed (**2**) FPA implementations for the receiving FSO terminal (Fig. 1).

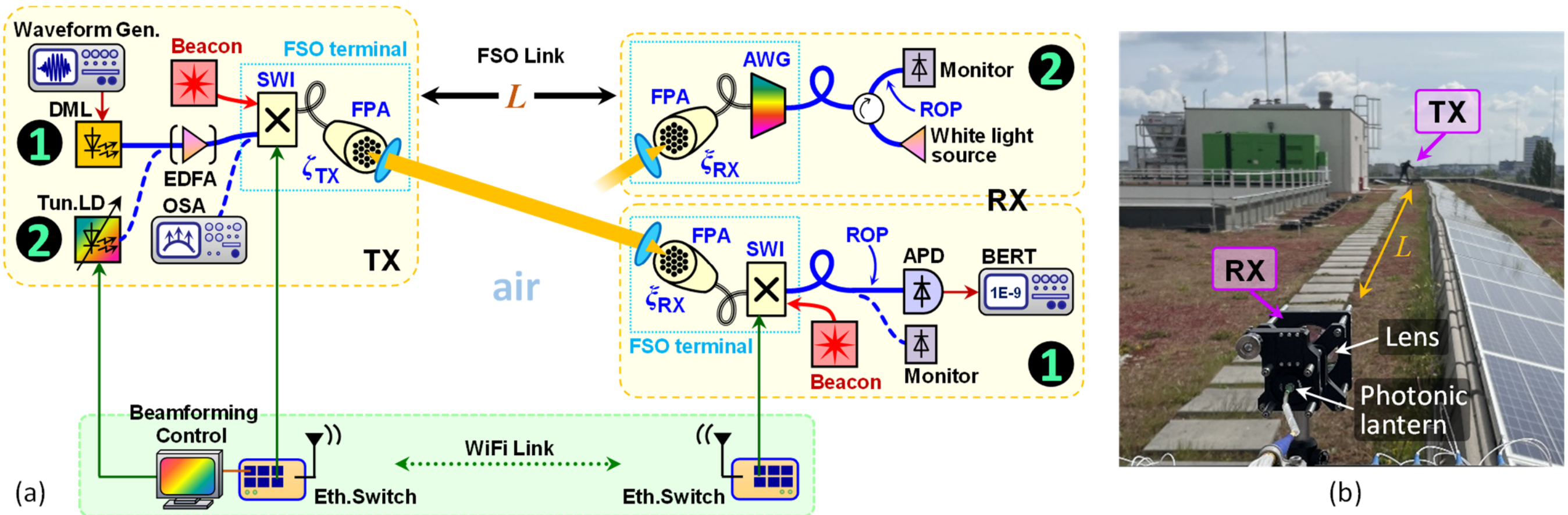

**Fig. 3.** (a) Experimental setup for FPA-based Fi-Wi-Fi bridge, and (b) out-door installation of a corresponding FSO link.

The space switch employed in this work provides an active beamforming network through its 1×*M* architecture based on micro electro-mechanic system (MEMS) technology, with *M* = {32, 64}. It provides a colorless solution that can be in principle operated across the entire spectral window of single-mode propagation over standard SMF, thus reaching from the O- to the L-band. This allows the implementation of the fiber-in-the-air paradigm for which FSO channels can transparently extend the audacity of fiber capacity.

For the wavelength-routed beamforming network, we employed a 96-channel arrayed waveguide grating (AWG) with a grid spacing of 50 GHz. Moreover, the AWG had a cyclic transmission with a free spectral range (FSR) of 5.96 THz. Since the AWG is a passive component, the selection of the antenna element can be simply made through tuning the emission wavelength of the data signal. This means that the FPA configuration is colored rather than spectrally transparent. It is therefore not compatible with optical wideband transmission. On the other hand, it does not require active actuation at the receiving FSO terminal, thus allowing for a more centralized beamforming control.

Figure 2 reports the typical losses associated to these two beamformer implementations. The average insertion loss for the photonic lantern, space switch and wavelength-routing AWG is 0.12, 1.06 and 5.76 dB, respectively, without attributing for connector losses. The accumulated insertion loss for both implementations is reported in Fig. 2(b) for all 61 antenna elements. Three of the elements ($\delta$) faced a high loss of more than 60 dB. This is attributed to a broken photonic lantern, as evidenced through the high loss in the corresponding insertion-loss histogram in Fig. 2(a).

## EXPERIMENTAL SETUP OF FPA-ASSISTED FSO LINK

The experimental setup is presented in Fig. 3(a). FPA beamformers are used at the transmitting (TX) and receiving (RX) FSO terminals as fiber-to-air interfaces. While at a transmitter a space-switched beamforming network is being employed, we evaluated a space-switched (**1**) and λ-routed (**2**) configuration for the receiving FSO terminal.

For scenario **1**, we feed the FPA beamformer with a directly modulated laser (DML). The DML operated at 1550 nm and had an output power of 10 dBm. The antenna element $\zeta_{TX}$ at the transmitting FPA is set through a 1×32 MEMS switch that is specifically connected to the inner 32 cores of the hexagonally arranged antenna element array. The light emerging at this focal plane is then collimated by an optical lens in order to yield a pencil beam that can be launched over the FSO link. At the receiving site, identical FPA optics are being used, whereas the beamforming network is comprised by a 64×1 MEMS switch. We monitored the received optical power (ROP) to analyse the coupling efficiency after initial pointing of the FPAs, which is facilitated through red beacon lasers. For data transmission, the DML is driven by an arbitrary waveform generator, while the received signal is detected by an avalanche photodetector (APD) with a sensitivity of -28.7 dBm at a BER of $10^{-10}$. The bit error ratio (BER) was evaluated by means of real-time error counting.

Scenario **2** differs primarily with respect to the AWG that is being used for antenna element selection at the receiving FPA beamformer. This further necessitates the use of a tuneable laser at the transmitter site in combination with the wavelength-selective choice for the receiving antenna element $\xi_{RX}$.

The diameter *D* and focal length *f* of the optical lens of the FPA have been chosen depending on the installation of this Fi-Wi-Fi bridge. For the preliminary in-door characterization, they were *D* = 6 mm and *f* = 10 mm. In case of the out-door FSO link that spans over a reach of *L* = 63 m between the transmitting and receiving FSO terminals, as shown in Fig. 3(b), these lens parameters were *D* = 2" and *f* = 100 mm. This out-door setting results in a beam diameter of ~28 mm and a discretization in step size of ~23 mm when steering the beam. An Erbium-doped fiber amplifier (EDFA) at the transmitting FSO terminal is additionally employed for the out-door link to compensate for the trunk fiber losses between the roof-top FSO link and our lab facilities, which hosted the data source and sink together with several opto-electronic transceivers.

Control of the opto-electronic beamforming elements is facilitated through an auxiliary management channel that has been implemented through a co-existing wireless radio channel. In case of scenario **2**, this control channel simplifies in a way that the selection of the receiving antenna element $\xi_{RX}$ associated to the passive, AWG-based beamforming network can be remotely made at the transmitting site. This is accomplished by simply tuning the optical seed source of the FSO link rather than actuating an active FPA component at the site of the FSO receiver.

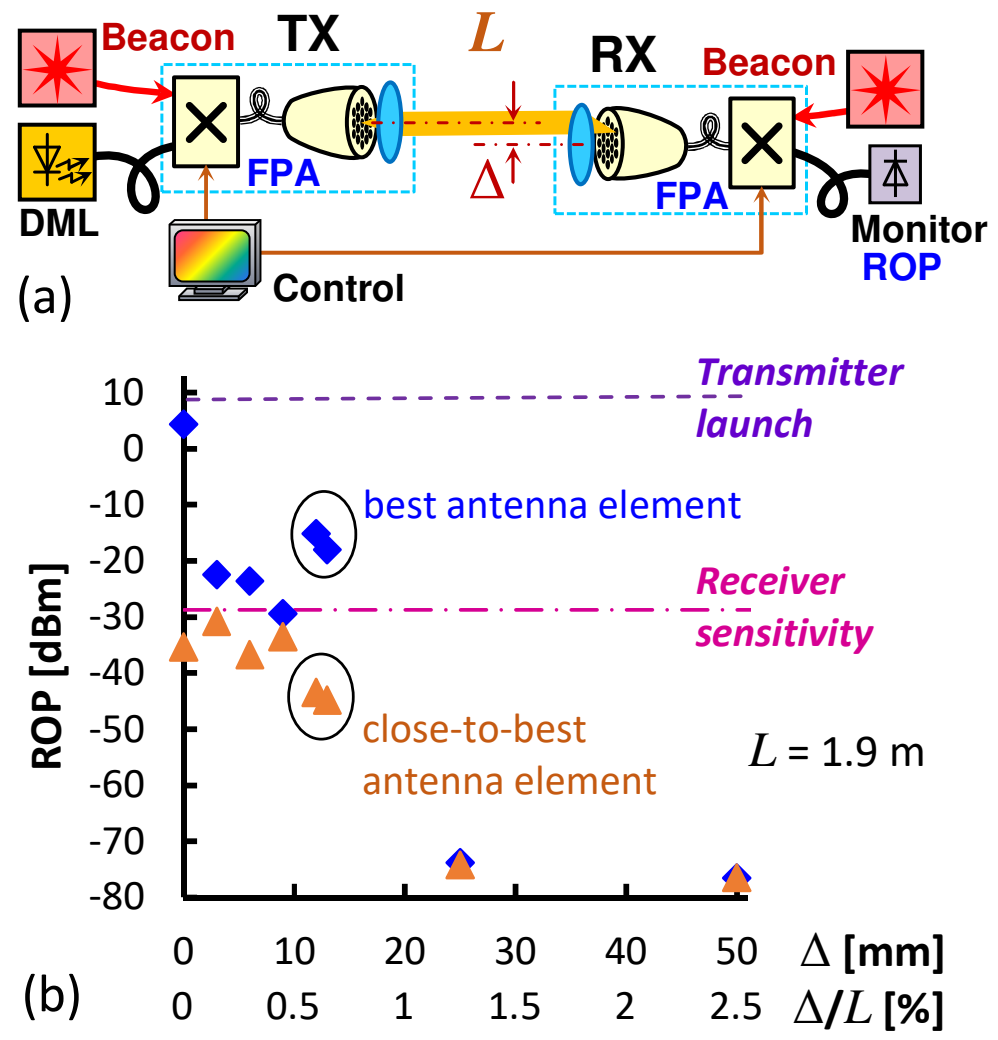

**Fig. 4.** (a) Characterization setup for FPA optics optimization. (b) Alignment tolerance to a lateral beam offset Δ.

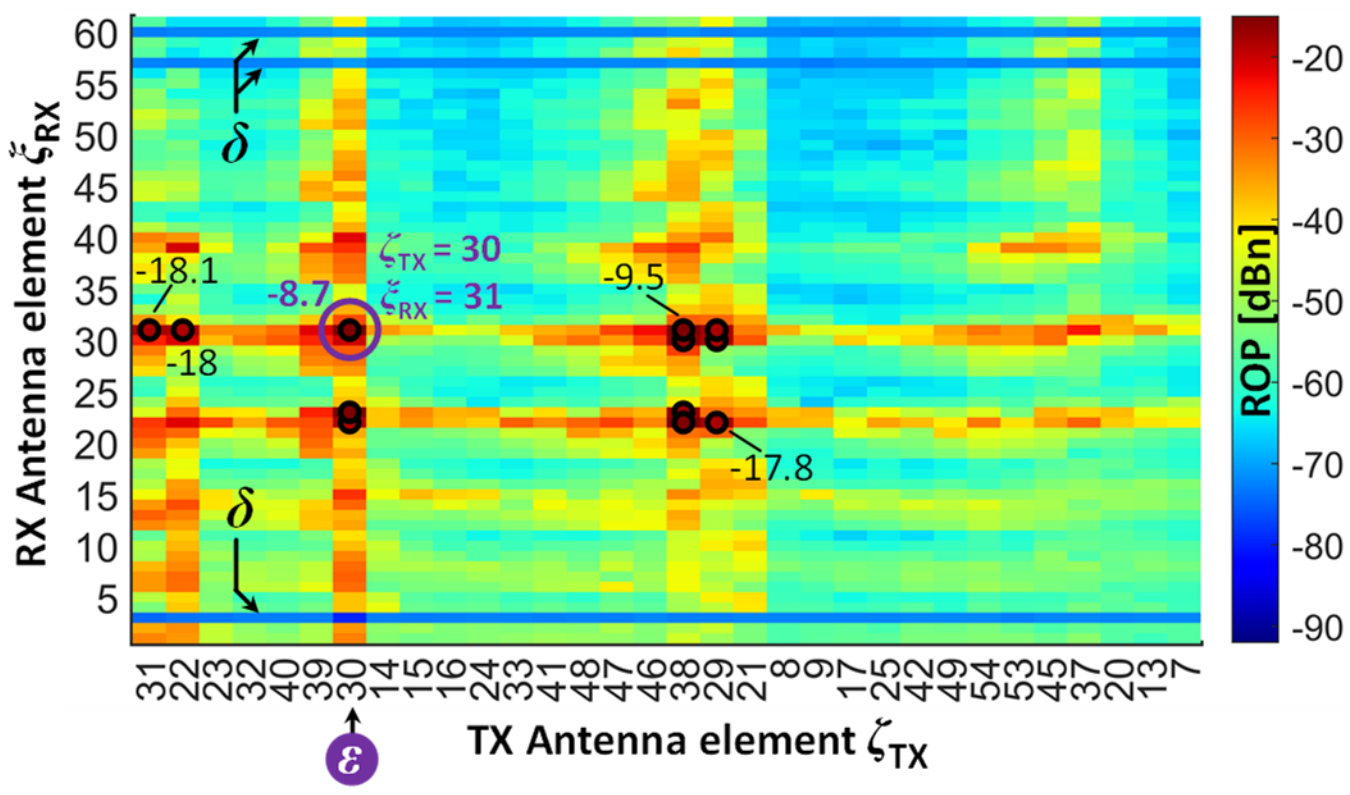

**Fig. 5.** Map of the coupled optical power for all 32×61 antenna element pairs ($\zeta_{TX}$, $\xi_{RX}$) of the space-switched FPAs after fine alignment through beamforming control.

## SPACE-SWITCHED FPA BEAMFORMER

### *A. Tolerable Beam Offset for Discretized Steering*

We first evaluated the beam-steering functionality of the space-switched FPA beamformer (scenario **1**) in the characterization setup shown in Fig. 4(a), which spans over reduced lab distance of $L$ = 1.9 m between both FSO terminals. The power injected into the FSO transmitter was 9.2 dBm. As a performance benchmark for this in-door FSO link we consider the very low coupling loss of 1.6 dB for the optimal face-to-face coupling between the central transmitter and receiver antenna elements (31 → 31), with additional losses of 1.6 dB incurring per MEMS switch at each of the FPA beamformers.

When the receiving FSO terminal is laterally offset by Δ from the initial face-to-face orientation, as indicated in Fig. 4(a), we noticed a strong dependence of the coupling efficiency on the exact offset value. This is effect is reported in Fig. 4(b) and is attributed to the short focal length of the lens, which determines the discretization $\upsilon$ of ~7 mm when steering the collimated beam at the given link distance through advancing to an adjacent FPA antenna element. This also explains the drop of 26.8 dB in ROP for Δ = 3 mm (◆), since there is no ideal antenna element available for collecting the transmitted optical signal with the given beam waist diameter of 2.8 mm. For an increased offset Δ, the ROP decreases slightly until a local maximum of -15 dBm (◆) is reached at an offset of Δ = 12 mm, or Δ/$L$ = 0.63% when expressed in terms of FSO reach. For this lateral offset, the decrease in ROP at the second-best antenna element (▲) confirms the improved coupling conditions at the prime antenna element. The coupling drops significantly for larger lateral offset values since the number of antenna elements is not large enough to support a wider field-of-view for the given design parameters of the FPA.

This initial characterization emphasizes the criticality of a mismatch between a small beam waist diameter and the discretization in beam steering for our FPA-enabled FSO links. For a given number of antenna elements, which yields the parameter $N$ as discussed in Section II, the lens parameters and thus the beam waist diameter have to be optimized according to the fixed antenna element pitch $\Pi$ at the focal plane, which together with the focal length $f$ defines the discretization $\upsilon$ in beam steering and therefore the maximum offset Δ that can be recovered under misaligned conditions. Given our scope on short-range terrestrial FSO links, such as required to establish a local Fi-Wi-Fi bridge, the maximum offset that needs to be accounted for after coarse initial pointing of the FSO terminals is clearly less critical than the discretization in beam steering, which also influences the minimum achievable coupling loss. In addition, we would like to mention that the angle-of-arrival of the incident beam plays a critical role for efficient SMF coupling. While we did not investigate this effect in detail, we believe that it can be neglected for short-range point-to-point FSO links due to the good overlap in beam spots – as discussed in the following – as well as the large number of antenna elements and thereby possible beam paths.

### *B. Out-Door FSO Link as Fi-Wi-Fi Bridge*

We then set up our FSO link at the roof-top location shown in Fig. 3(b). This out-door link bridges a gap of $L$ = 63 m between two SMF ports. Given this larger distance, the ratio of beam diameter to step size is critical and needs optimization, since the beam diameter of 2.8 mm was less than half of the step size of 7 mm for the in-door setup. This provides a better FPA setup since the individual beam spots impinging upon the receiving FPA do now overlap by ~32% due to the longer focal length of $f$ = 100 mm.

The link was installed by performing a coarse sequential pointing of the FSO terminals towards each other, using a red beacon laser and visual alignment. After this coarse initial pointing procedure, we launched an EDFA-boosted laser emission at 20 dBm into the transmitting FSO terminal to ensure a good contrast in ROP for the acquisition to follow, and performed a finer alignment by means of beamforming control, which elaborates the optimal antenna element pair(s) using the ROP as a link quality indicator.

The link calibration results after this alignment mechanism are presented in Fig. 5 as ROP map for all pairs ($\zeta_{TX} \rightarrow \xi_{RX}$) of antenna elements. Considering settling times for these ROP measurements, the alignment procedure took about 5 minutes for our non-optimized alignment process. However, due to the typical MEMS switching times of 10 ms, the whole 32×61 sweep could also be performed in ~30 seconds. Out of all pairs, two favourable combinations can be identified: 30 → 31 and 38 → 31. These pairs yield a ROP of -8.7 and -9.5 dBm, respectively. Expressed as fiber-to-fiber loss, these ROP values would correspond to a FSO link loss of 28.7 and 29.5 dB, respectively. The three broken lantern ports $\xi_{RX}$ = {3, 57, 60} show no transmission ($\delta$ in Figs. 2 and 5).

We further conducted a continuous acquisition of the ROP for the optimal pair of antenna elements. Results are shown in Fig. 6(a), together with the actual weather data during the 8-hours recording on a cold Central European day in April. Stable optical coupling between the SMF ports at the transmitting and receiving FSO sites has been maintained without the need for repeated beam steering. Even though the coupling appears to be not directly impacted by a change in ambient temperature, which showed a swing of 4°C over the entire acquisition period, we noticed a momentary thermal impact due to the temporary exposure to solar influx. This is evidenced in Fig. 6(a) through the peaking in solar irradiance for sporadic clear-sky conditions during noon ($\beta$) and in the late afternoon, which heats up the opto-mechanic elements involved at the transmitting and receiving FSO terminals. We attribute the correlated (positive) change in optical coupling of up to 2 dB to this thermal perturbation. The three short bursts of +3 dB in ROP ($\alpha$) are attributed to sudden wind gusts. The histogram for the ROP, presented in Fig. 6(b), shows a probability peak at -8.2 dBm for the ROP distribution and does not indicate deep fading of the coupled optical power. The ROP finds itself within an interval of 3.2 dB during 90% of the measurement time, between -10.9 and -7.7 dBm.

We then activated data transmission through 10-Gb/s on-off keyed modulation of the DML current. The data signal was fed to the roof-top FSO link through 360 m of vertical trunk fiber cable, where it is boosted by the EDFA. After pencil-beam transmission over the out-door Fi-Wi-Fi bridge, the signal was relayed back to the lab premises through another 200 m of trunk fiber, where it is received by the APD receiver.

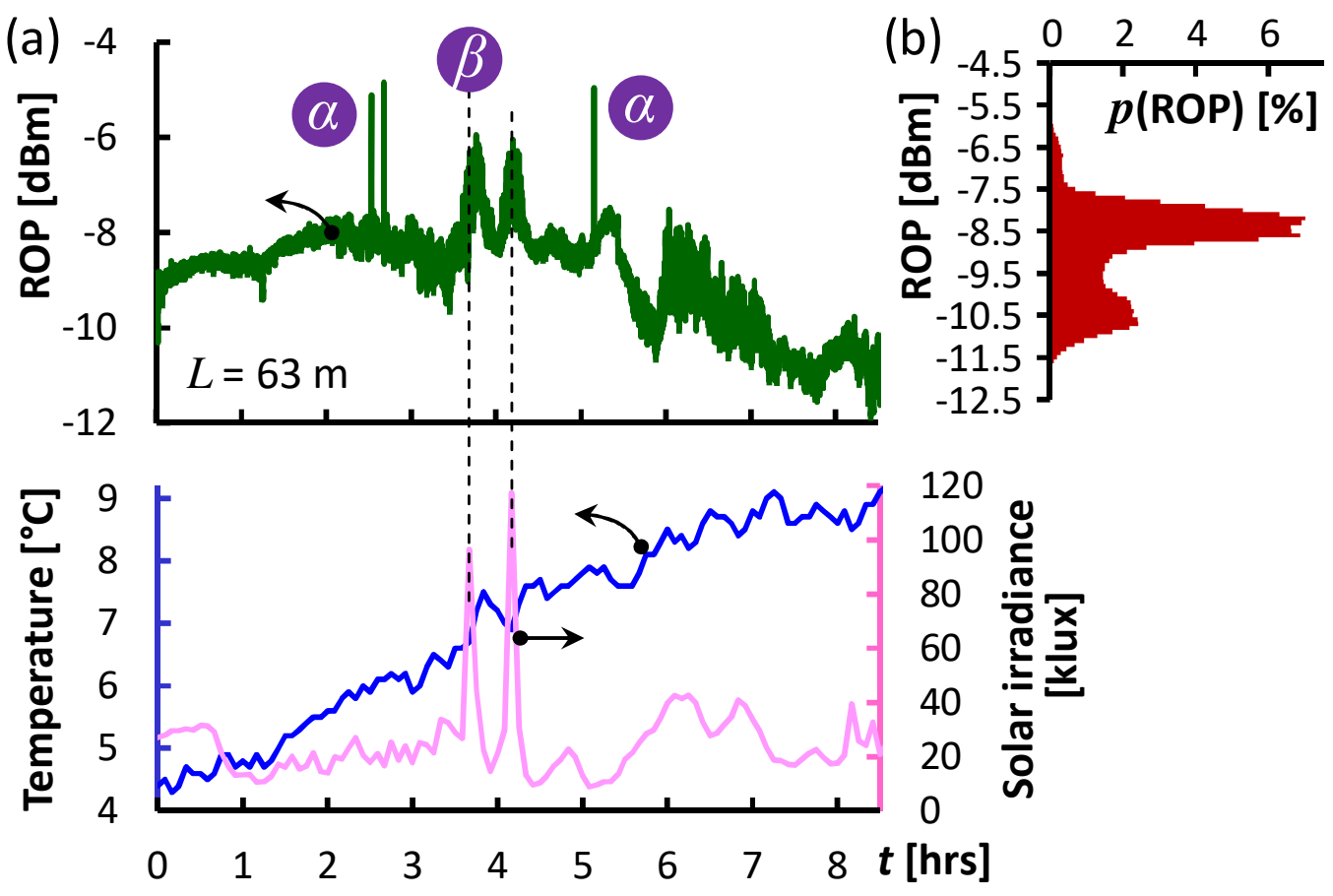

**Fig. 6.** Long-term FSO link stability for an out-door Fi-Wi-Fi bridge and weather data for the acquisition period.

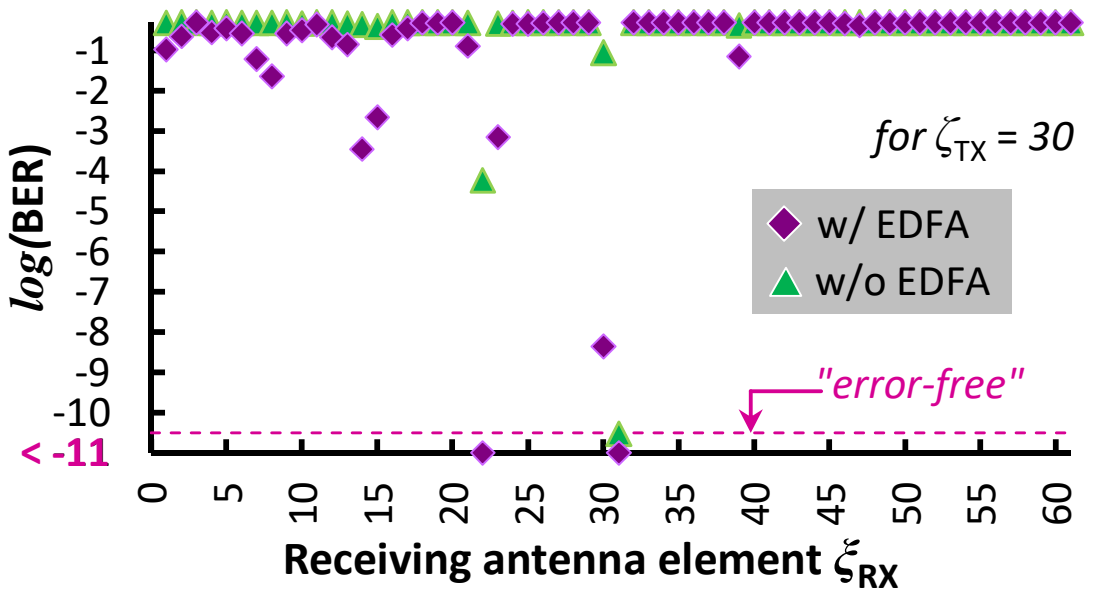

**Fig. 7.** BER performance for the amplified and unamplified out-door Fi-Wi-Fi bridge for data signal reception over all receiving antenna elements.

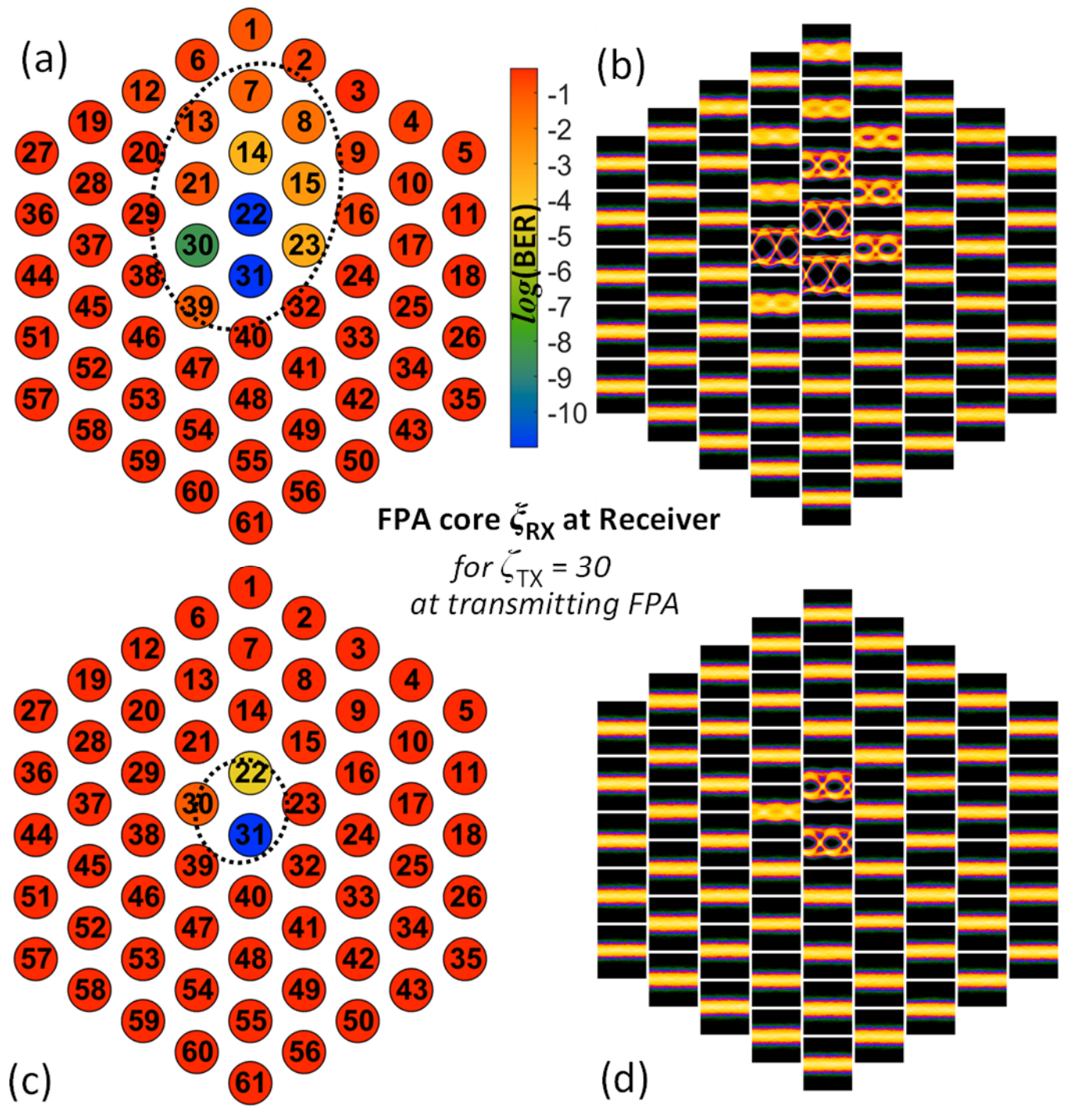

**Fig. 8.** Spatial BER distribution over the focal plane of the receiving FPA, and corresponding 10-Gb/s eye diagrams with (a, b) and without (c, d) EDFA booster at the transmitter.

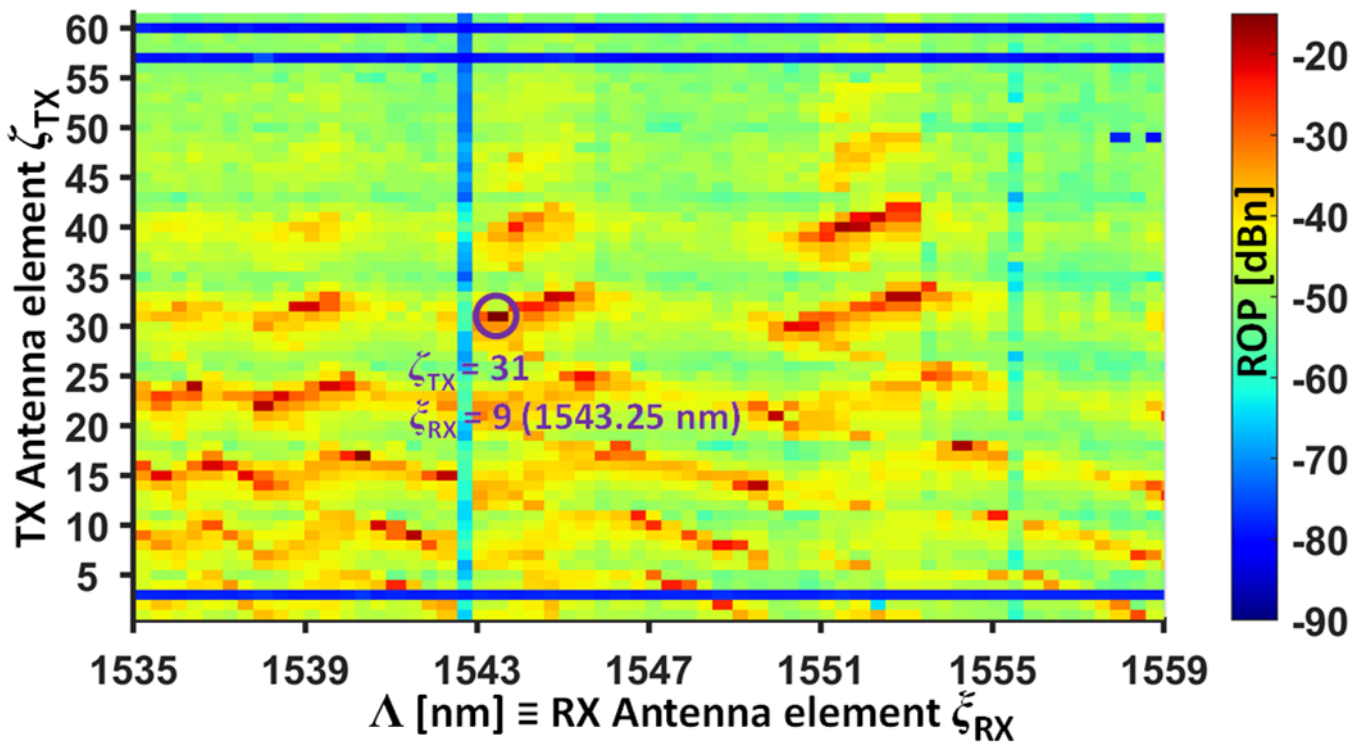

**Fig. 9.** Map of the coupled optical power for all 61×61 antenna element pairs ($\zeta_{TX}$, $\xi_{RX}$) of the hybrid space-switched (TX) and wavelength-routed (RX) FPAs after fine alignment through beamforming control.

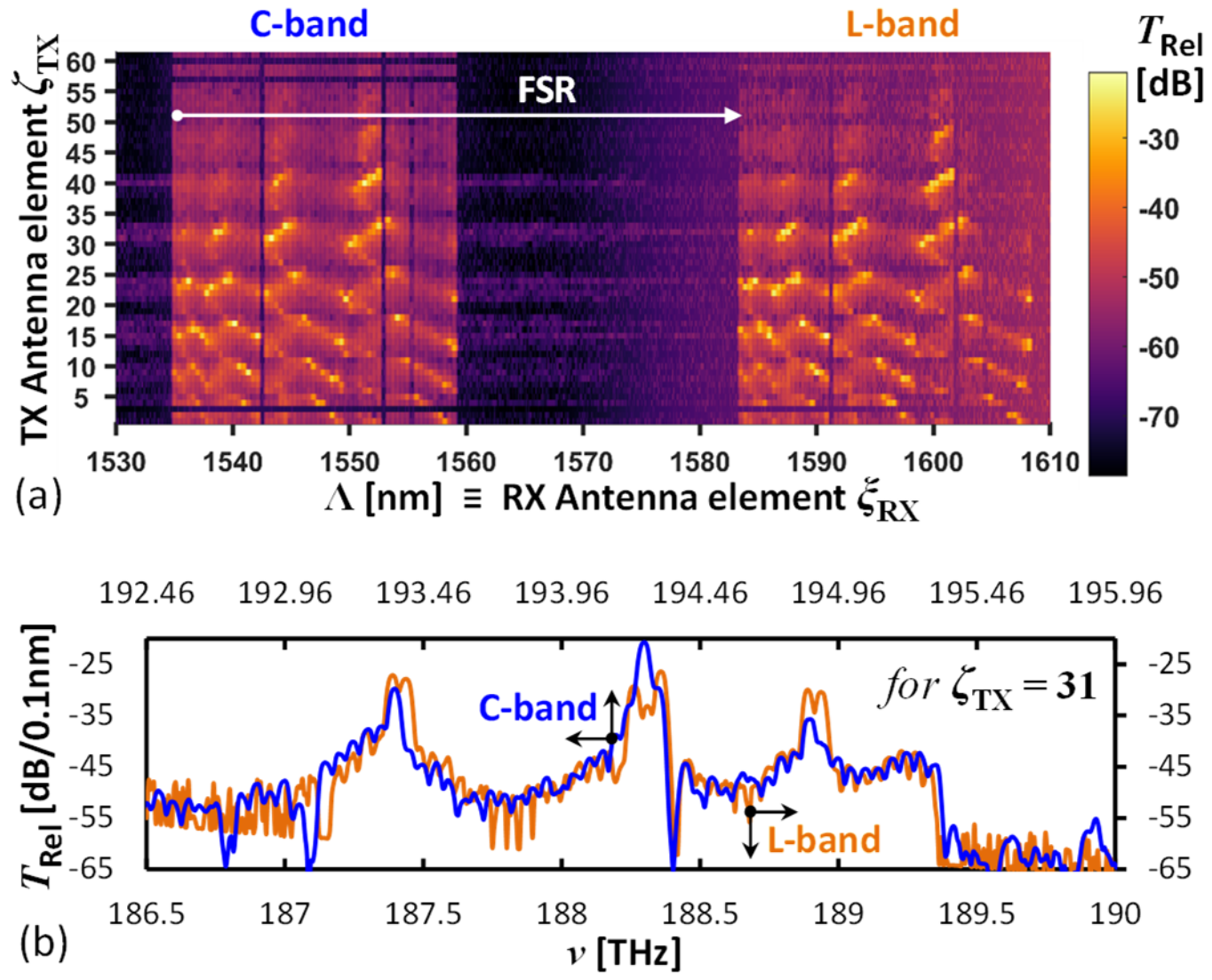

**Fig. 10.** Simultaneous channel sounding in an adjacent waveband and overlap of C- and L-band signatures after white-light transmission.

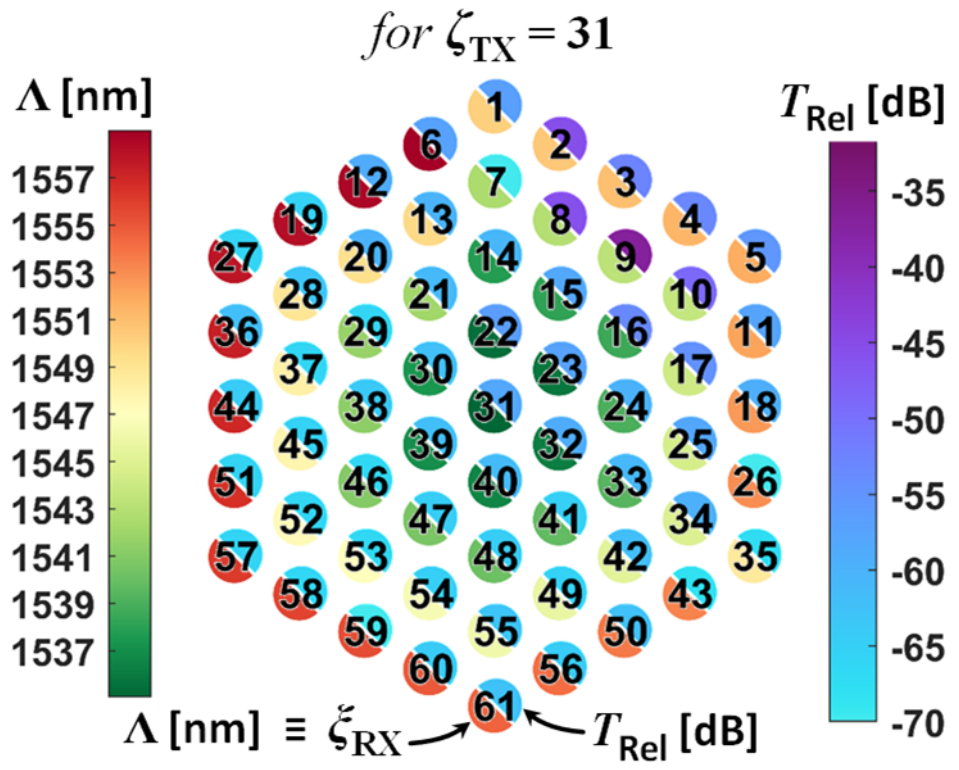

**Fig. 11.** Spatial power distribution over the focal plane of the receiving FPA, together with the assigned wavelength channel.

According to the optimal coupling among the FSO terminals, we chose a fixed FPA antenna element $\zeta_{TX}$ = 30 at the FSO transmitter ($\varepsilon$ in Fig. 5) while optimizing the antenna element $\xi_{RX}$ at the receiving FPA in order to investigate a change in the field-of-view. Given the larger beam spot, there are multiple pairs $\zeta_{TX} \rightarrow \xi_{RX}$ of antenna elements that yield a ROP that surpasses the sensitivity of the APD receiver for the amplified FSO link. Out of the 18 combinations that feature a power margin of more than 6 dB before the relay over the receiver-side trunk fiber, those within a margin of 10 dB to the optimal configuration are exemplarily highlighted in Fig. 5. In terms of BER, we achieve values below $10^{-11}$ at $\xi_{RX}$ = 31 (ROP of -14 dBm) and at $\xi_{RX}$ = 22 (ROP of -23.8 dBm), as summarized in Fig. 7 (◆) over all antenna elements $\xi_{RX}$. The graphical representation of the BER over the focal plane of the receiving FPA, shown in Fig. 8(a), indicates an elliptic shape in the BER distribution and the corresponding eye diagrams (Fig. 8(b)). This distribution covers more than five antenna elements and indicates a slight angular pointing error between the transmitting and receiving FSO terminals. This can be expected since the initial pointing was purely based on a visual alignment of visible beacons. In addition, we believe that a slight offset from the ideal focus at the receiving terminal reduced the chance of the incident beam being focused on an area without suitable fiber core, which is motivated by the low fill-factor, yet at the expense of additional loss. It nonetheless proves that a large power margin of at least 9.8 dB in ROP can be accomplished towards the sensitivity level of the APD receiver. When removing the EDFA booster before the FSO transmitter, we can still achieve data transmission over the unamplified Fi-Wi-Fi bridge: We accomplish a BER below $10^{-11}$ at $\xi_{RX}$ = 31 (Fig. 7, ▲), which yields a ROP just above -28 dBm. The corresponding BER map and eye diagrams are presented in Fig. 8(c, d).

### Wavelength-Routed FPA Beamformer

In case that ultra-wideband operation is not a primary objective, the FPA beamformer can rely on a wavelength-routed beamforming network (scenario **2**). We have evaluated such a configuration by replacing the lower-radix switch with an AWG as described in Section II, while re-using the higher-radix switch in the transmitting FPA beamformer. This yields a configuration with 61×61 antenna pair combinations in total, which has been evaluated in an in-door setting over a distance of $L$ = 6 m, yet using the same optics as employed for our out-door FSO link.

The corresponding coupling among all pairs of FPA antenna elements is presented in Fig. 9, whereas the antenna elements at the receiving FSO terminal are now labelled in terms of their AWG-defined C-band wavelength. This definition that is made through the physical connection between photonic lantern and AWG follows a spiral-shaped inwards-out allocation, meaning that the center of the focal plane marks the lowest wavelength of 1535.06 nm. As shown in Fig. 9, the launch at the transmitting FPA antenna element $\zeta_{TX}$ = 31 results in a maximum for the optical coupling at λ = 1543.25 nm, which translates to the receiving antenna element $\xi_{RX}$ = 9. As for the space-switched FPA architecture, there are multiple pairs with a similarly good coupling, owing to the characteristics of the beam propagation involving two FPA beamformers.

At first glance, this result suggests that there is no obvious advantage gained when replacing the space switch of the beamforming network with the AWG, given its increased insertion loss (Fig. 2) and the reduction in continuous optical bandwidth that is provided for the FSO channel. However, the wavelength-routed FPA architecture offers an important feature as it enables simultaneous channel sounding on an adjacent spectral waveband, by virtue of its cyclic transmission function. This is evidenced in Fig. 10(a), which presents the received optical spectrum at the transmitting FSO terminal relative to the transmitted optical power for the case that a C+L-band white-light source at 13 dBm is counter-injected into the FSO link at the site of the receiver (Fig. 3(a)). The signal waveband in the C-band region from 1535.06 to 1558.93 nm is spectrally replicated in the L-band, so that the information obtained through spectral monitoring of the L-band transmission of the injected white light can be used to determine the optimal signal wavelength at the C-band, and thus the optimal pair ($\zeta_{TX}$, $\xi_{RX}$) for the FPAs of the FSO link. Figure 10(b) shows this exemplarily for $\zeta_{TX}$ = 31 through the corresponding overlap in sounded spectra in the C- and L-bands, which features the very similar spectral signature at an optical frequency $v$ that is shifted by the FSR of the AWG. Figure 11 then performs a translation of this information that has been acquired during channel sounding exclusively in the L-band ($T_{Rel}$) in order to determine the respective C-band signal wavelength ($\Lambda$), given also the aforementioned spiral-shaped wavelength assignment at the focal plane of the receiving FPA. This proves how simultaneous channel sounding can potentially enable continuous monitoring and optimization of the beamformer setting, without interrupting data transmission in an adjacent waveband. It shall be stressed that the L-band power of 10 dBm applied for channel sounding allows the link to be operated in an eye-safe regime.

## Conclusion

We have experimentally demonstrated an optical FPA beamformer that acts as an air interface for an FSO link between two SMF ports. The architecture of the FPA has been closely aligned to SMF-based components that are employed as its constituent elements, involving well-known fiber-optic components such as optical switches or AWGs as beamforming networks, whereas the multi-element antenna configuration at the focal plane is based on a photonic lantern. Both, a space-switched and a wavelength-routed beamforming network have been evaluated, each of them contributing through distinctive advantages such as wide optical transparency or simultaneous channel sounding. Stable SMF-to-SMF coupling has been accomplished after coarse initial pointing through involving beamforming control, as demonstrated over a space-switched out-door Fi-Wi-Fi bridge for error-free 10 Gb/s single-wavelength data transmission at a large power margin of 10 dB. Bonding multiple antenna elements for the purpose of multi-beam FSO transmission or the mitigation of atmospheric turbulence is left for future work.

***Acknowledgement:*** *This work has received funding from the Smart Networks and Services Joint Undertaking (SNS JU) under the European Union's Horizon-Europe research and innovation programme under Grant Agreement No. 101139182.*